# A Communication Model for Adaptive Service Provisioning in Hybrid Wireless Networks

MATTHIAS R. BRUST AND STEFFEN ROTHKUGEL
Faculty of Sciences, Technology and Communication
University of Luxembourg
L-1359 Luxembourg
Luxembourg
Mr_Brust@yahoo.de, Steffen.Rothkugel@uni.lu

*Abstract:* - Mobile entities with wireless links are able to form a mobile ad-hoc network. Such an infrastructureless network does not have to be administrated. However, self-organizing principles have to be applied to deal with upcoming problems, e.g. information dissemination. These kinds of problems are not easy to tackle, requiring complex algorithms. Moreover, the usefulness of pure ad-hoc networks is arguably limited. Hence, enthusiasm for mobile ad-hoc networks, which could eliminate the need for any fixed infrastructure, has been damped. The goal is to overcome the limitations of pure ad-hoc networks by augmenting them with instant Internet access, e.g. via integration of UMTS respectively GSM links. However, this raises multiple questions at the technical as well as the organizational level. Motivated by characteristics of small-world networks that describe an efficient network even without central or organized design, this paper proposes to combine mobile ad-hoc networks and infrastructured networks to form hybrid wireless networks. One main objective is to investigate how this approach can reduce the costs of a permanent backbone link and providing in the same way the benefits of useful information from Internet connectivity or service providers. For the purpose of bridging between the different types of networks, an adequate middleware service is the focus of our investigation. This paper shows our first steps forward to this middleware by introducing the Injection Communication paradigm as principal concept.

## 1 Introduction

In recent years, mobile devices as cellular phones, notebooks, MP3-players, digital cameras, etc. became more commonplace in our daily life. In addition to *being mobile*, the desire of *being connected* with other devices or, in particular, the Internet, arises immediately. Technological advances like Wi-Fi and Bluetooth for local ad-hoc communication as well as UMTS and GSM for linking to a backbone network drive progress in this respect.

Based on the transmission medium, networks can be classified in wired and wireless networks. Commonly, wireless networks are divided into *infrastructured* and *infrastructureless* (ad-hoc) networks. Examples for *infrastructured* networks are cellular networks, wireless ATM, Mobile IP, and satellite networks. *Infrastructured* networks might offer access points (base stations) for the wireless integration with a fixed wired network backbone. Mobile ad-hoc networks in turn are purely wireless, volatile in nature, and without access to a backbone.
In a lot of situations, an integration of both technologies would be an asset. To illustrate that, in the subsequent section, the disadvantages of both "pure" *infrastructured* and *infrastructureless* approaches are discussed.

The applicability and usefulness of pure ad-hoc networks is generally limited, due to the volatile nature of the environment. Devices are mobile, entering and leaving each others coverage area frequently. The density of devices is also subject to change. If the density drops below a certain critical threshold, communication might even become impossible. Moreover, mobile devices are resource-constrained in terms of computing power, storage capabilities, as well as power supply, and are potentially unreliable. Depending on the technology, low bandwidth might be an additional disadvantage of ad-hoc networks, for instance when using Bluetooth. Hence, finally no guarantees can be given to the users of such pure ad-hoc networks.

A different aspect can be motivated as follows: devices constituting a mobile ad-hoc network normally act also as routers, thus allowing to establish so-called multi-hop ad-hoc networks. The big advantage is that there is no need for central administration and no additional costs for being connected, due to the lack of a network provider.

Global administration, however, has to be substituted by self-organization, which is a hard problem.

The following example is taken from the domain of public transportation. Systems as the one described hereafter exist in reality in different cities. Terminals connected to an *infrastructured* network are installed e.g. on bus stations, offering an information service for passengers. Here, GPS information is evaluated to estimate the bus' arrival time. A fixed display shows the resulting information at the bus station. The installation of weather-resistance and energy-consuming hardware induces substantial costs for the provider. The physical presence of that expensive hardware makes it vulnerable to vandalism. Moreover, information can only be provided at fixed locations, i.e. directly at bus stops. In case of persons being in doubt of getting the bus in time, this pure *infrastructured* solution appears as not really attractive.

Hybrid solutions integrating infrastructured with ad-hoc networks might provide considerable improvements. Due the use of mobile devices as service hosts, the service can generally be provided anywhere, increasing the quality of the service significantly. Furthermore, eliminating the need for the information terminals reduces costs for the service provider drastically. However, the latter nevertheless also results in a major drawback. Communication between service provider and mobile hosts using UMTS or GSM causes additional costs that normally the *clients* have to pay. Thus, reducing costs in that sense is a concrete goal.

Besides technical issues, costs, and energy consumption, there is another reason that mobile entities might not use the backbone link directly. In the case that information has been generated locally, for privacy reasons and for communication efficiency, the services on each mobile device might be interested to keep group-related data within the group only, rather than to share group-sensitive data across the whole network by using the network backbone.

Certainly, pure infrastructured settings aim at situations where users normally do not have access to any mobile device. The tendency, however, is that the use and accessibility of mobile devices providing services in the described manner is significantly increasing. Thus, for the near future the proposed combination between infrastructure and ad-hoc networking is going to become more and more relevant.

This work focuses on hybrid wireless networks to cope with the limitations of both infrastructured and ad-hoc networks, combining their advantages in a symbiotic and synergic way. For this, mobile devices are considered of being able to link to UMTS or GMS to connect to a network backbone and additionally to be capable of communicating in an ad-hoc fashion with nearby devices using e.g. Bluetooth or Wi-Fi. A couple of questions need to be tackled, like for instance: how is it possible to create hybrid wireless networks providing Internet connectivity at low costs and energy consumption? What theoretical and practical restrictions do hybrid wireless networks obey? Since there are some concrete hints that the designed hybrid network environment can be compared with small-world networks with a small characteristic path length, the approach seems to be promising.

For the purpose of bridging between the different types of networks, an adequate middleware service is the focus of our investigation. This paper shows our first steps forward to design and implement this middleware. For an efficient integration, a communication paradigm called *Injection Communication* is introduced and discussed in detail in the subsequent sections.

## 2 Injection Communication

As discussed in the previous section, both ad-hoc networks and infrastructured networks have their advantages and disadvantages. The envisioned hybrid wireless networks offer benefits as instant service provisioning respectively Internet connectivity to basically every device in the network. Multi-hop paths can extend the coverage area of access points and augment mobility at low cost and low energy consumption. Several issues have to be investigated both at the technical as well as the organization level. The communication in such a hybrid network is proposed to be *injection-driven* and, thus, significantly different from the client/server paradigm to take the technical differences into account. Injection Communication is introduced in the following example, discussing its diverse aspects.

Suppose a physical group or *clique* composed of several nearby persons being interested in certain information or in using a specific service. Referring again to the public transporting system example, on an *infrastructured* wireless network a service is installed that evaluates GPS information to estimate the arrival time of buses. In contrast to the setting described initially, there is no fixed display showing the resulting information at the bus station. However, the service can be used on any mobile device that is registered for this service and thus can be used anywhere. It is further supposed that there are some people waiting at the bus stop.

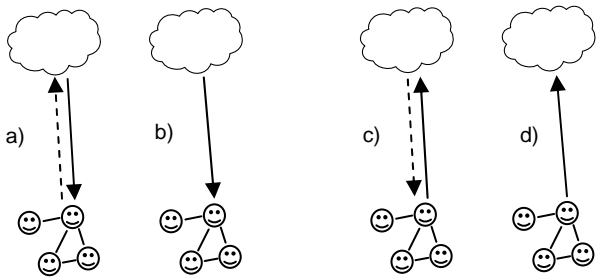

*Fig.1:* Backbone Injection (left) and Clique Injection (right).

Some members of the clique are interested in the bus' arrival time in general, and delays to be expected in particular. Thus, there is a shared interest in using this service and several ways to get the desired information. An initial attempt could be to first ask neighbors within the ad-hoc network if there is current information already available. If this is not the case, information must be requested explicitly by establishing a cost-intensive backbone link (see Fig. 1a).

It should be considered, however, that the mobile devices of passengers at the bus stop may communicate in an ad-hoc fashion basically at no costs. Hence, in the scenario mentioned, it is an efficient way to select one adequate device that will exclusively request the information of interest by using a backbone link. The service sends the information to that device which is acting as gateway. In a subsequent step, the information can be distributed to other devices within the ad-hoc network in an epidemic way. Resulting costs may also be distributed to all other devices that declared their interest in that particular information.

### 2.1 Backbone Injection
The process described is called *Backbone Injection*, considering that information is being injected from the backbone to one device being member of the ad-hoc network. That device in turn further disseminates the information to other interested parties, thereby overall minimizing costs and maximizing benefits. Due to the temporary nature of the backbone link, information inconsistencies appear that need to be dealt with.

In Backbone Injection it is possible to explicitly register for, respectively request, injections by the clique or by just one member of the clique (see Fig. 1b). The envisioned middleware also covers cases that don't deal with joining groups to get information, but instead allowing the individual decision of a single mobile entity to update information. This entity-driven injection can also be used to distribute information in an ad-hoc fashion.

### 2.2 Clique Injection
Information can also be actively created on devices residing within cliques. Observing learning groups whose members are using their mobile devices to develop an idea, some results might be interesting for other learning groups as well as for the tutor for evaluation reasons. The backbone might potentially be interested in such information. Thus, cliques are also able to initiate injections, called *Clique Injection*. Like Backbone Injections, Clique Injection can also be spontaneous (see Fig. 1d) as well forced (see Fig. 1c) by the backbone. Spontaneous Clique Injection is used when e.g. the learning group has done a considerable progress in their work. The backbone may force or request a Clique Injection, when it needs updated information e.g. for forwarding to other learning groups.

### 2.3 Wormhole Injection (Tunneling)
Another aspect is covered by so-called *Wormhole Injections*. A Wormhole Injection uses the availability of the backbone to tunnel information though it to another clique located somewhere else; thus introducing inter-clique communication. The Wormhole Injection concept is a logical step in introducing Injection Communication in hybrid wireless networks.

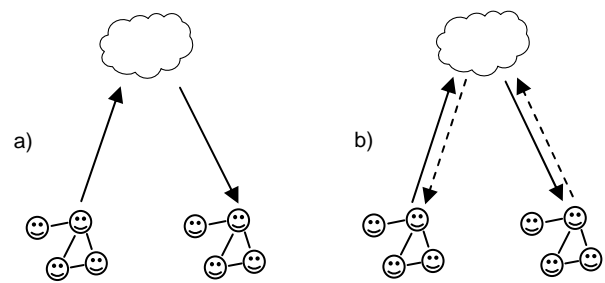

*Fig.2:* Wormhole Injection without (a) and with (b) components on backbone.

Wormhole Injection may occur in two ways. In the first way, the backbone network serves just for information forwarding from one clique to the other. No software component is required to be located inside the backbone. From a logical point of view, this also opens the possibility to abandon all entities located inside the backbone, and provide a fully distributed infrastructure without any centralization. Due to this fact, cliques are acting in a fully self-organized fashion, but in contrary to traditional mobile ad-hoc networks, with an additional dimension for communication (see Fig 2a). In some cases it might be necessary for one clique to request information on the backbone not being interested what clique is delivering the information. For this, the backbone must have a service that initiates a Clique Injection for gathering the requested information. This process is transparent to the requested clique (see Fig. 2b). The advantage of this approach is that information does not have to be stored and, so, administrated on the backbone.

## 3 Injection Process

The envisioned middleware has to evaluate parameters to plan and manage an injection. One parameter deals with the injection initiator that could be the service being located on a mobile device as well as the service on the backbone. In the Backbone Injection case, a well-suited device of the clique should be identified serving as an *Injection Point*. That device has to control and optimize epidemic distribution of information within the target ad-hoc network. This injection process is described more detailed in this section.

### 3.1 Injection Points

Different important criteria for electing the Injection Point exist. Consideration includes available power, technical equipment, load balancing issues, and for instance also the time a device is expected to remain available. An example for the latter criterion is that it is known which bus a passenger intends to take and hence when he is supposed to leave the group. Furthermore, the current environment and relationship to its neighbors is important for one clique member. Supposing that the device is high clustered and so one of the central members of a group, epidemic behavior for information spreading will take effect faster.

Because of constraints related to information inconsistencies or time, the backbone may force a backbone injection to synchronize its set of information. In this case, a backbone injection is initiated spontaneously as described in Section 2.

Device registration is necessary because a spontaneous backbone injection can only be done if a concrete Injection Point device is known. As in Sun et al. [3], it is assumed that devices are required to register for a certain service on the backbone or that devices register autonomously when entering in a marked geographical area by using GPS information and a register service. Figure 1 illustrates the method proposed for operating spontaneous backbone injection.

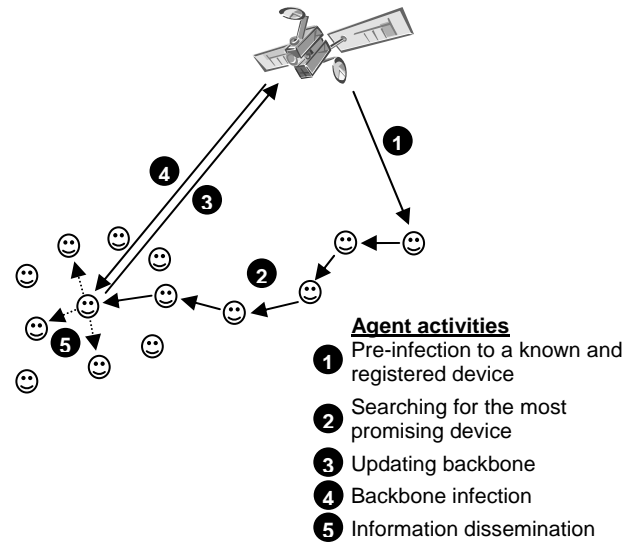

*Fig.3:* Creating an Injection Point.

A spontaneous backbone injection is initiated by contacting a registered device. But for the backbone injection to be handled efficiently, a proper mobile device has to be selected that adheres to the aforementioned criteria. A similar problem for discovering this optimal or semi-optimal device is already known in a different context, namely how to discover marketplaces as described in Hutter et al. [9]. Adaptive mobile agents are proposed to tackle this problem as it is described in Section 4.

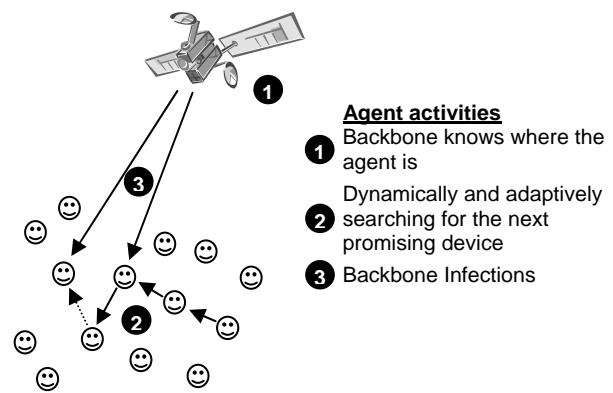

*Fig.4:* Selecting an adequate Injection Point.

Fig. 4 shows mobile code searching for an adequate Injection Point. Over time, however, the topology might change. For example, because a bus arrived, several people leave while others join the group at the bus stop. The Injection Point has to react accordingly to the changing network structure. Based on its predictions and on the current network structure, the hosting device has probably to be changed for re-fulfilling the selection criteria. The

Injection Point creation described in Fig. 3 does not need to be executed each time to minimize communication traffic.

## 3.2 Applying several Injection Points

Due to mobility and fast topology changes, a sparse ad-hoc network or multiple network partitions cause problems in efficiently spreading information. It would even be efficient to discover groups of interests in an ad-hoc network to efficiently establish one backbone link to each group. For creating logical groups, user profiles or active applications could be used. Thus, strategies of infecting several mobile entities at the same time have to be considered. Obviously, this is a challenging problem requiring proper heuristics and adaptive algorithms.

# 4 Inconsistence of Information

Due to the temporary nature of the links in the envisioned environment, information inconsistencies between mobile devices and backbone network appear. Thus, the applications we are targeting are supposed not to depend on full consistency of the information provided, but to explicitly allow some degree of fuzziness instead. Maintaining fully consistent views on the data on all devices would require an immense effort in terms of communication and synchronization and thus make the proposed approach unfeasible. For plenty of different application domains, however, it can well be assumed that a limited degree of consistency in fact is sufficient, but in compensation the service is practical for clients in economical terms and in terms of the availability of the service. As one example, the public transportation system shows that there is an individual invariant as the maximum time for waiting before asking for clients to use that service, so that the need for service connecting appears only if this invariant is injured. In that case a backbone injection may be initiated. In terms of information inconsistencies, the important challenge is to find out when to do a backbone injection by also considering these individual invariants.

In the following sub-section terms such as *Information Item*, *Consistency Properties*, etc. are introduced to deal with information inconsistencies between mobile entities and backbone network.

## 4.1 Information Items, Providers, Seekers

The data an application is working on is structured as *Information Items*. Information Items can be considered as uniquely identifiable chunks of data, obeying certain semantics. Examples are the arrival time of a specific bus at a particular stop, information about a certain product, and individual questions in an M-Learning scenario.

Information Items are being made available to others through *Information Providers*. In case some data relevant for others is being produced by a particular device, the device itself might play the role of an Information Provider, making the data available for others. The data might also be forwarded to a different device, which then acts as a delegate Information Provider. Applications interested in querying data in turn are called *Information Seekers*. Each device may represent an Information Provider, an Information Seeker, or both.

## 4.2 Consistency of Information Items

Due to the nature of the envisaged environment, it is sensible to allow multiple copies of an information item to co-exist at the same time at different locations. It is assumed furthermore that these copies do not necessarily need to be kept fully in sync with each other. This is feasible due to the fact that many applications do not require total information consistency, but allow a certain degree of information fuzziness instead.

A description of the consistency characteristics and requirements are attached to each particular information item in the form of C*onsistency Properties*. These properties are driving the process of synchronizing multiple copies of a single information item that are maintained by several Information Providers. The propagation of updates is controlled and optimized by taking the Consistency Properties into account. Aside from that, other factors determined by the relationships and link types between the set of information providers need to be considered as well.

As mentioned above, information items are considered to be active entities, propagating themselves among multiple information providers. However, this process can be augmented by application-driven strategies further influencing the information dissemination. Each user is enabled to contribute actively by defining *Consistency Requirements* in an individual way. The visibility of the Consistency Requirements, however, can be limited in scope. They are relevant only for the Information Providers serving their originators. Finally, Consistency Properties in conjunction with Consistency Requirements drive the information dissemination process.

As one concrete example, the Consistency Requirements in the bus example might be stricter for a business traveler than for a tourist.

## 5 Related Work

Service provisioning in mobile ad-hoc networks is investigated in Handorean et al. [5]. Maamar et al. [8] expand the pure mobile ad-hoc model and include to their considerations fixed entities that provides resources or services. This approach is motivated because of the objective to make composite Web Services available in the targeted network. As pointed out in [6] and [7], composite Services serves as a main problem for mobile ad-hoc based middleware. Sun et al. [3] proposed hybrid wireless networks to combat the limitations of *infrastructured* wireless networks and provide Internet connectivity to *ad hoc* networks. Two routing schemas are introduced to deal with different application requirements. Sun et al. [3], also Ratanchandani et al. [10] focus on network with Mobile IP capability (with foreign agents etc.) and Internet Gateways to communicate with wired correspondent nodes. As result, it is possible to deduce that Mobile IP and on-demand routing protocols in a mobile ad-hoc network can work together to set up multi-hop paths to a foreign agent in the network, allowing Internet connectivity [10]. Andreadis [2] uses a fixed Internet Gateway as for an access point to provide Internet connectivity to the entire ad-hoc network.

## 6 Conclusion

Due to their characteristics, ad-hoc networks can not fully satisfy applications that require a certain QoS. This problem is tackled by introducing a backbone network link that is intended to be used on demand only, driven by application requirements in order to obtain a QoS level required.

General problems inherent to this approach include, but are not limited to, dealing with information inconsistencies that occur due to the nature of the Injection Communication between backbone network and mobile entities. Creating and maintaining Injection Points may be another important issue as well as creating a cost and payment model for the envisaged environment. An adequate cost model is a basic aspect influencing the Injection Implementation, but payments in ad-hoc networks are difficult to handle and more investigation has to be made to deal with related problems.

The next step is to develop an adaptive multi-agent architecture at the implementation level. This approach seems to be promising for realizing middleware to foster applications in the realm of hybrid wireless networks.

**Acknowledgements**
This research is supported in parts by the Luxembourg Ministère de la Culture, de l'Enseignement Supérieur et de la Recherche.